\documentclass[a4paper]{jpconf}
\usepackage{graphicx} % Required for inserting images
\usepackage{iopams}
\usepackage{amsmath}
\usepackage[hidelinks]{hyperref}
\usepackage[noabbrev]{cleveref}
\hypersetup{
  pdftitle={Structure of 13C in the framework of the Cluster Shell Model},
  pdfauthor={A.H. Santana Valdés, Roelof Bijker},
  pdfsubject={Nuclear Structure},
  pdfkeywords={cluster, cluster shell model, form factors},
  pdfinfo={Journal=Journal of Physics Conference Series,
  Event=44SNP23,
  Year=2023}
 }
\newcommand{\bal}{\begin{align}}
\newcommand{\eal}{\end{align}}

\begin{document}
\title{Structure of $^{13}\mathrm{C}$ in the Cluster Shell Model}
\author{A H Santana Valdés$^1$, R Bijker$^2$}
\address{Instituto de Ciencias Nucleares, Universidad Nacional Autónoma de México, Apartado Postal 70-543, 04510 Cd. de M\'exico, M\'exico}
\ead{$^1$adrian.santana@correo.nucleares.unam.mx, $^2$bijker@nucleares.unam.mx}

\begin{abstract}
We study the structure of 13C in the framework of the Cluster Shell Model. A comparison of the available experimental data with our model is made. Some predictions for level ordering and form factors are presented.
\end{abstract}

\section{Introduction}

Recent measurements of rotational excitations in $^{12}$C have renewed interest in the cluster structure of light nuclei, particularly $\alpha$ conjugate nuclei as clusters of $k$ $\alpha$-particles. The interest in these structures dates as back as 1937, with the works of \cite{wheeler,Teller,Brink1,Brink2}. The intricate patterns in which protons and neutrons are arranged give rise to a multitude of special properties, many of them of great implications, from nuclear physics to astrophysics, where the best example of this is the Hoyle state. Clustering in nuclei is complex, and the underlying physics is not yet fully understood, being a major challenge for researchers. Over the years $^{12}$C has been the object of study of several models, including Antisymmetric Molecular Dynamics\cite{AMD}, Fermion Molecular Dynamics\cite{FMD}, BEC-like cluster model\cite{BEC}, lattice EFT\cite{EFT1} and the Algebraic Cluster Model\cite{ACM2}, all of them implementing or obtaining the cluster structure as a result. A more comprehensive review of many different models and their assumptions about clustering in nuclei can be found in \cite{FreerFynbo,Schuck,Freer}.

Yet for its neighbors, a query arises when one questions to what extent the cluster structure persists with adding extra nucleons. This proceeding addresses part of this question, showing precursory results of longitudinal form factors for $^{12}$C and $^{13}$C and transverse form factors for $^{13}$C in the framework of the Cluster Shell Model (CSM), where $^{13}$C is seen as a $^{12}$C core plus an extra neutron. $^{13}$C can then be considered as a system with ${\cal D}'_{3h}$ symmetry, consisting of three $\alpha$-particles in a triangular configuration plus an additional neutron moving in the deformed field generated by the cluster. The matrix elements of the form factor operators are then calculated, where a dominance of the core cluster structure is found for longitudinal form factors, exhibiting a very similar $q$ dependence between $^{12}$C and $^{13}$C, corroborated by experimental data. Finally, we show some preliminary results for transverse form factors.

\section{Cluster Shell Model}
The Cluster Shell Model was introduced in \cite{CSM1,CSM2,PPNP} to describe nuclei composed of $k$ $\alpha$-particles plus additional nucleons, denoted as $k \alpha + x$ nuclei. It is similar in spirit to the Nilsson model \cite{Nilsson}; the main difference lies in that the odd nucleon moves in the deformed field generated by the cluster core. 
 \subsection{Cluster Potential}
 To obtain the single-particle energy levels, we define
\begin{align}
H \;=\; T + V(\vec{r}) + V_{\rm so}(\vec{r}) + \frac{1}{2}(1+\tau_3) V_{\rm C}(\vec{r}) ~,
\label{hcsm}
\end{align}
{\it i.e.} the sum of the kinetic energy, a central potential obtained by convoluting the density 
\begin{flalign}
\label{rhor}
\rho(\vec{r})=&\left( \frac{\alpha}{\pi}\right)^{3/2} 
\sum_{i=1}^{k}\exp \left[ -\alpha \left( \vec{r}-\vec{r}_{i}\right)^{2}\right] 
\nonumber \\
=& \left( \frac{\alpha }{\pi }\right)^{3/2} \mbox{e}^{-\alpha(r^{2}+\beta ^{2})} 
\, 4\pi \,\sum_{\lambda\mu} i_{\lambda}(2\alpha \beta r)\,Y_{\lambda\mu}(\theta,\phi)\sum_{i=1}^{k} Y_{\lambda\mu}^{\ast}(\theta_{i},\phi_{i}) ~,  
\end{flalign}
with the interaction between the $\alpha$-particle and the nucleon, a spin-orbit interaction, and a Coulomb potential for an odd proton. In \cref{hcsm}, $\vec{r}_i=(r_i,\theta_i,\phi_i)$ represent the coordinates of the $\alpha$-particles with respect to the center-of-mass of the cluster structure. The case of $^{13}$C belongs to $k=3$ with an equilateral triangular configuration, whose coordinates are given by $(\beta,\frac{\pi}{2},0)$,  $(\beta,\frac{\pi}{2},\frac{2\pi}{3})$ and $(\beta,\frac{\pi}{2},\frac{4\pi}{3})$. The deformation parameter (from spherical symmetry) is $\beta$, which is the distance of each of the alpha particles to the center of mass of the cluster structure.
The resulting potentials are 
\begin{align}
\label{V0}V\left(\vec{r}\right)&=-V_{0}\sum_{\lambda\mu}f_{\lambda}\left(r\right)Y_{\lambda\mu}\left(\theta,\phi\right)\sum_{i=1}^{k}Y_{\lambda\mu}^{*}\left(\theta_{i},\phi_{i}\right),\\
 \label{Vs0}V_{so}\left(\vec{r}\right)&=V_{0,so}\frac{1}{2}\left[\frac{1}{r}\frac{\partial V\left(\vec{r}\right)}{\partial r}\left(\vec{s}\cdot\vec{l}\right)+\left(\vec{s}\cdot\vec{l}\right)\frac{1}{r}\frac{\partial V\left(\vec{r}\right)}{\partial r}\right],\\
\label{flambda} f_{\lambda}\left(r\right)&=e^{-\alpha\left(r^{2}+\beta^{2}\right)}4\pi i_{\lambda}\left(2\alpha\beta r\right).
\end{align}

\subsection{${\cal D}'_{3h}$ symmetry}
 
 \begin{figure}[h]
\centering
\includegraphics[width=14pc]{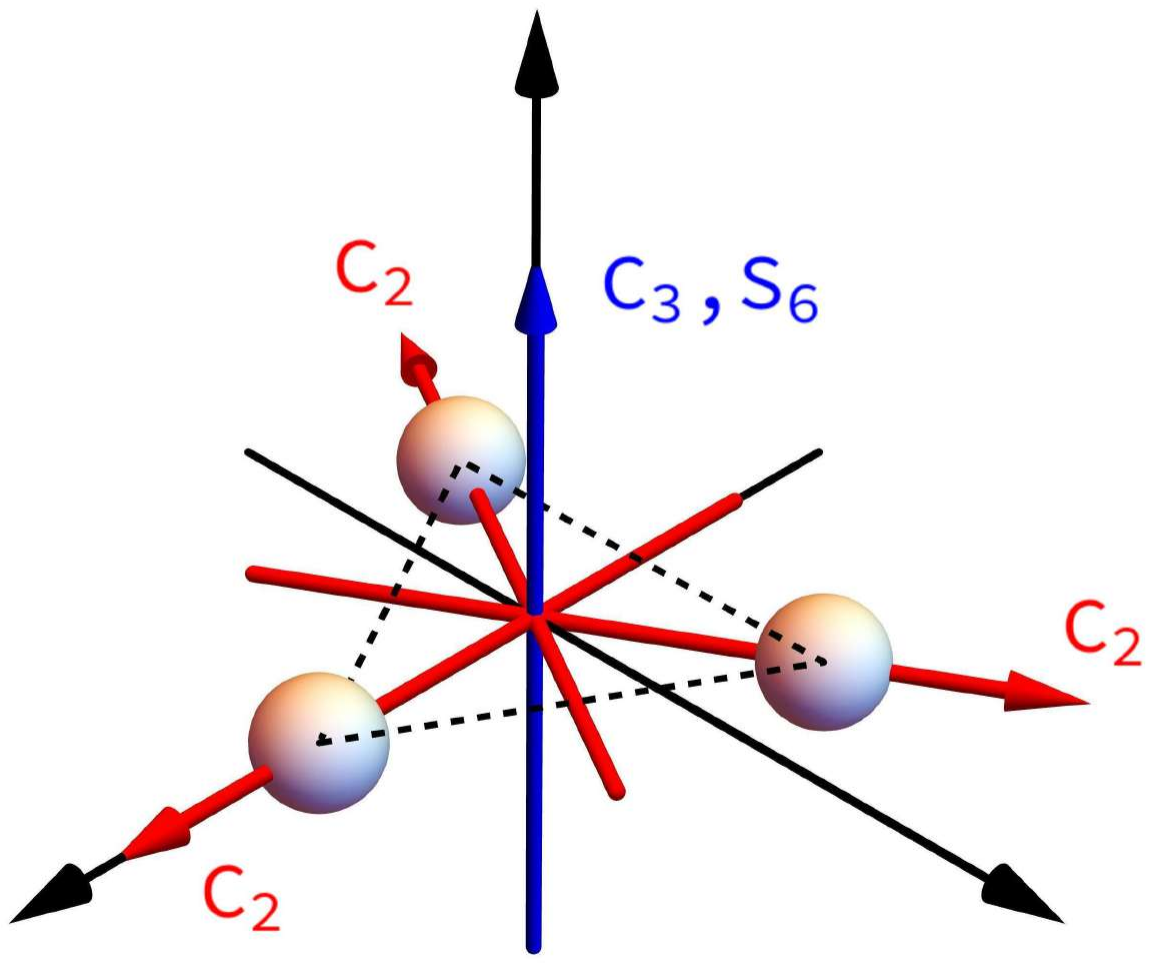}
\caption{Schematic view of all principal axis and rotation operators for ${\cal D}'_{3h}$ symmetry.}
\label{allaxisC13}
\end{figure}

 To construct a complete wave function we need to understand in more detail the ${\cal D}'_{3h}$ symmetry. 
 The objective is to construct a symmetry-adapted basis for ${\cal D}'_{3h}$ symmetry instead of a spherical basis\cite{bijker5,adrian}. In the case of triangular symmetry the eigenstates of \cref{hcsm} can be classified according to the doubly degenerate spinor representations of the double point group ${\cal D}'_{3h}$ \cite{Herzberg3,koster}: $\Omega=E_{1/2}$, $E_{5/2}$ and $E_{3/2}$, or in the notation of Ref.~\cite{bijker5} $\Omega=E_{1/2}^{(+)}$, $E_{1/2}^{(-)}$ and $E_{3/2}$, respectively. From similar work in molecular physics, to differentiate between the degenerate states, one finds all rotations of the structure of study; the case in question is shown in \cref{allaxisC13}. Then improper rotations are used (in this case the $S_{6}$) to identify the intrinsic states of the doubly degenerate spinor representations. The resulting labels are $\gamma=\pm 5/2$ for $\Omega=E_{5/2}$, 
$\gamma=\pm 1/2$ for $E_{1/2}$ and $\gamma=\pm 3/2$ for $E_{3/2}$.
 The Hamiltonian of the CSM is solved in the body-fixed system, using the harmonic oscillator basis $| nljm \rangle$. 
\begin{align}
\phi_{\Omega\gamma} =& \sum_{nljm} C^{\Omega\gamma}_{nljm} \left| nljm \right>
\label{phiharmonic}
\end{align}
The rotational states can be labeled by the angular momentum $I$, parity $P$, and its projection $K$ on the symmetry axis, $| I^P K \rangle$. Both $I$ and $K$ are half integers. Then the allowed values of $K^P$ for each one of the spinor representations, along with their $\gamma$ labels are given by \cite{bijker5,adrian}
\begin{align}
\begin{array}{lcl}
\Omega=E_{1/2}, \;\gamma=\pm\frac{1}{2} &:& K^P = \pm\frac{1}{2}^+, \mp\frac{5}{2}^-, \pm\frac{7}{2}^-, \ldots \\ 
\Omega=E_{5/2}, \;\gamma=\pm\frac{5}{2} &:& K^P = \mp\frac{1}{2}^-, \pm\frac{5}{2}^+, \mp\frac{7}{2}^+, \ldots \\ 
\Omega=E_{3/2}, \;\gamma=\pm\frac{3}{2} &:& K^P = \pm\frac{3}{2}^+, \mp\frac{3}{2}^-, \mp\frac{9}{2}^+, \ldots 
\end{array}
\label{rotband}
\end{align}
with $I=|K|$, $|K|+1$, $\ldots$. A schematic view can be seen in \cref{RotD3hFix}.
The complete wave function is defined then by \cite{adrian}
\begin{align}
\left| \Omega\gamma;I^PMK \right> = \sqrt{\frac{2I+1}{16\pi^2}} \psi_v 
\left( 1 + {\cal S}_i^{-1} {\cal S}_e \right) \phi_{\Omega\gamma} D_{MK}^I(\omega)
\label{TotWaveF}
\end{align}
where $\psi_v$ is the vibrational wave function, $\phi_{\Omega\gamma}$ 
the intrinsic wave function and $D_{MK}^I(\omega)$ the rotational wave function. 
The wave function is invariant under the transformation ${\cal S}_i^{-1} {\cal S}_e$ 
where the operator ${\cal S}$ is the product of a rotation about $\pi$ followed by a parity transformation \cite{adrian}. 
The operator ${\cal S}_i^{-1}$ acts on the intrinsic wave function and ${\cal S}_e$ on the rotational wave function.

 \subsection{Splitting of single-particle levels}
\begin{figure}[h]
\centering
\includegraphics[width=22pc, height=18pc]{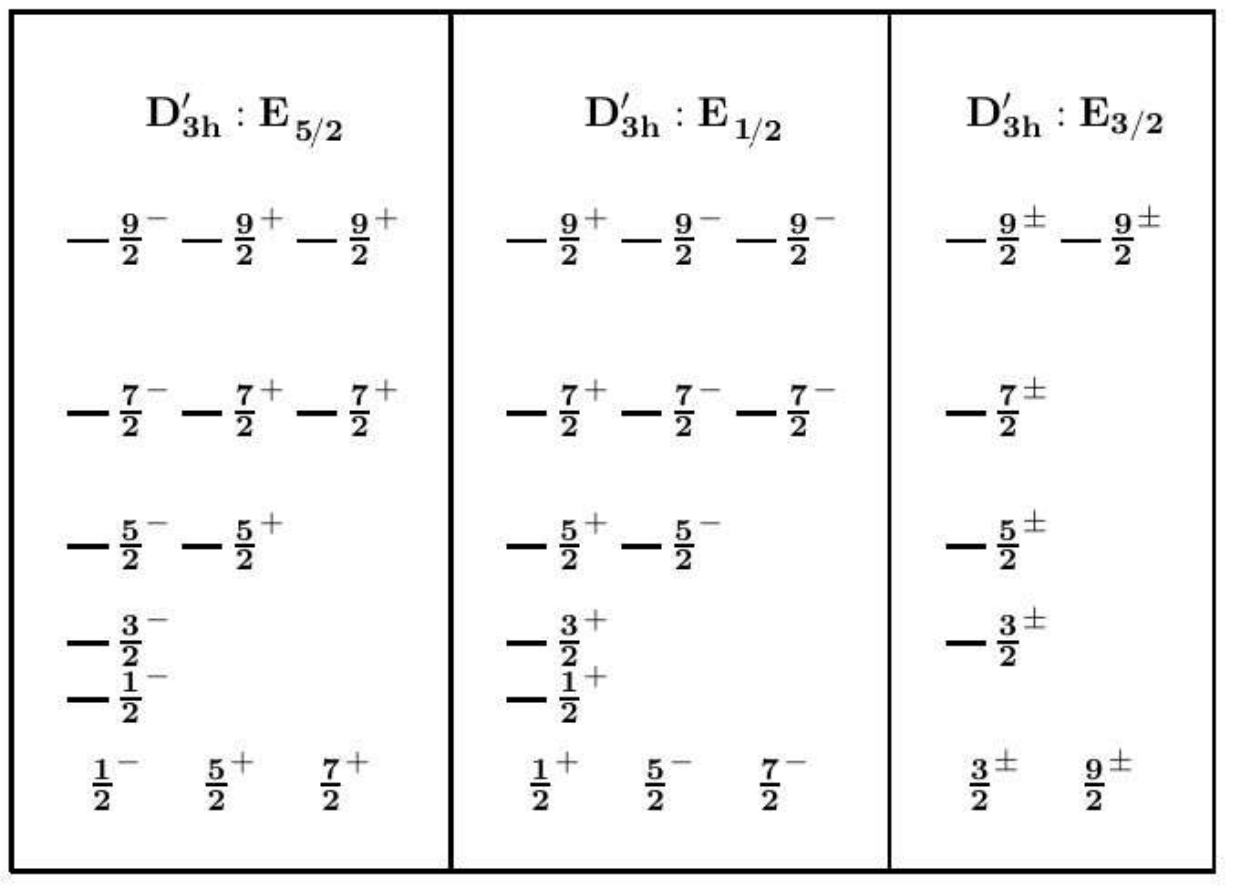}
\caption{\label{RotD3hFix}Schematic view of rotational structure with  ${\cal D}'_{3h} $, proposed for $^{13}$C. Imagine obtained from \cite{confadrian2}.}
\end{figure}

\begin{figure}[h]
\centering
\includegraphics[width=16pc,height=16pc]{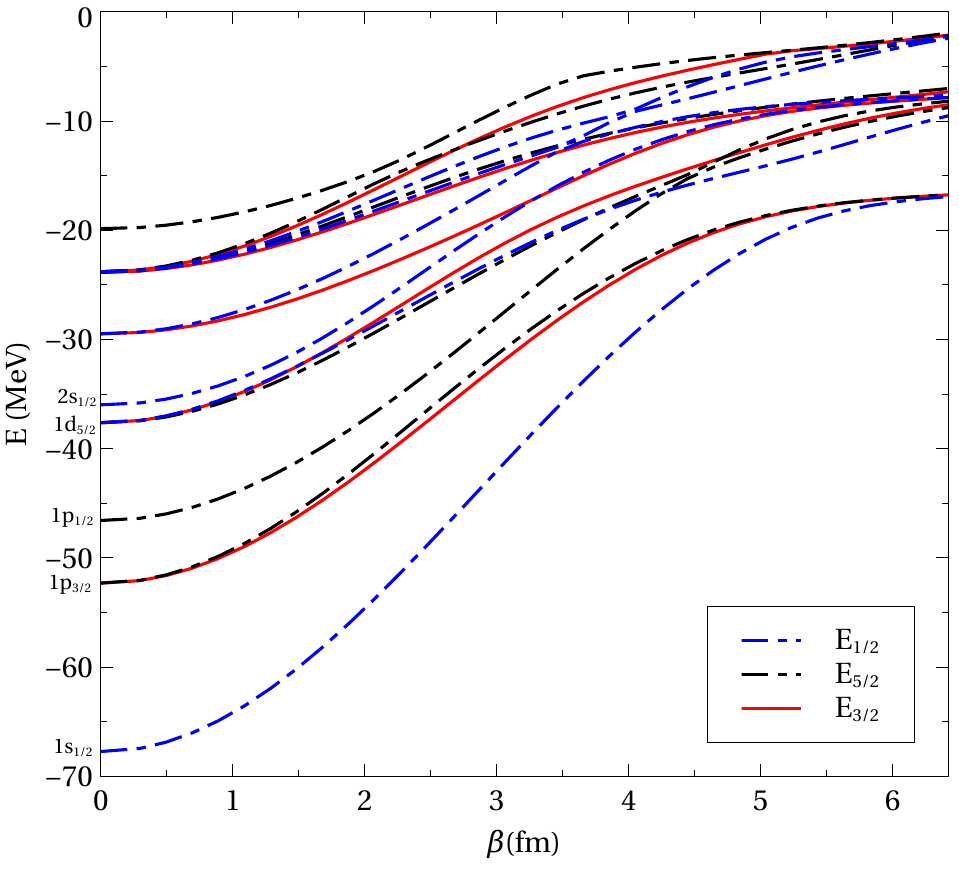}
\caption{Single-particle energy level splitting from solving \cref{hcsm}. The values used for \cref{V0,Vs0} are $V_{0}$=32 MeV, $\alpha$=0.0511 fm$^{-2}$ and $V_{0,so}$=17 MeV fm$^{-2}$.}\label{NivTrianFull}
\end{figure}

 Previous work on the splitting of single-particle levels is shown in \cite{CSM1,confadrian}. The splitting of the single-particle energy levels is seen in \cref{NivTrianFull}. Two simple observations that are illustrated in the figure are that at a value of $\beta=0$ the spherical symmetry is recovered, and at max value, we have 3-fold degeneracy of the spherical symmetry. In other words, at the max value we obtain the view of a particle in a field formed by three separate (not bound in a nucleus) alpha particles, which is unlike the Nilsson model where in the case of max deformation one obtains the splitting for an "infinite cigar". From the elastic form factors, one obtains the value for $\beta$ and then comes back to \cref{NivTrianFull,RotD3hFix} and obtains the base rotational band proposed for $^{13}$C. Foreshadowing the subsequent section, the value obtained for $\beta$ is 1.71 fm, which in turn gives the ground state rotational band as $\Omega=E_{5/2}$, and the ground state with $I^P = \frac{1}{2}^-$. 

\section{Form Factors and comparison to experimental data}
Using \cref{TotWaveF} we can now calculate the form factors for $^{13}$C and compare them with available experimental data. We use the standard form factors operators, of which a full explanation is found in \cite{TdF}.

\subsection{Longitudinal form factors}

The charge distribution is taken to be \cref{rhor}, plus a point-like distribution for the extra nucleon
\begin{align}
\rho (\vec{r}) = \frac{(Ze)_{\rm c}}{3} \left( \frac{\alpha }{\pi }\right)^{3/2} 
\sum_{i=1}^{3}\exp \left[ -\alpha \left( \vec{r}-\vec{r}_{i}\right)^{2} 
\right] + \tilde{e} \delta(\vec{r}-\vec{r}_{\rm sp}),
\label{EchargeTot}
\end{align}
$(Ze)_{\rm c}$ is the electric charge of the $3\alpha$ core nucleus, and $\tilde{e}$ 
the effective charge of the extra nucleon, which for the case of $^{13}$C is taken to be zero. Making a Fourier transformation of \cref{EchargeTot}, and then summing over final and averaged over initial states, we obtain the multipoles of the longitudinal (Coulomb) form factor;
\begin{align}
F_{C\lambda}\left(q;\Omega'\gamma',I'K'\rightarrow\Omega\gamma,IK\right)&=\frac{\sqrt{4\pi}}{\left(Ze\right)_{\mathrm{odd}}}\langle I',K',\lambda,K-K'\left|I,K\right\rangle\delta_{\Omega\Omega'}\delta_{\gamma\gamma'}G_{vv'}^{\mathrm{c}}\left(q\right),
\label{FFlambda}
\end{align}
where $(Ze)_{\rm odd}$ denotes the electric charge of the odd nucleus.
For the vibrationally elastic case with $v=v'=0$ the collective part is given by 
\begin{align}
G^{\rm c}_{00}(q;\lambda,K-K') &= (Ze)_{\rm c} \, j_\lambda(q\beta) \, \mbox{e}^{-q^2/4\alpha} \, Y_{\lambda,K-K'}(\frac{\pi}{2},0)\, \delta_{K-K',3\kappa}
\label{gcoll}
\end{align}

\begin{figure}[h]
\centering
\includegraphics[width=19pc,height=16pc]{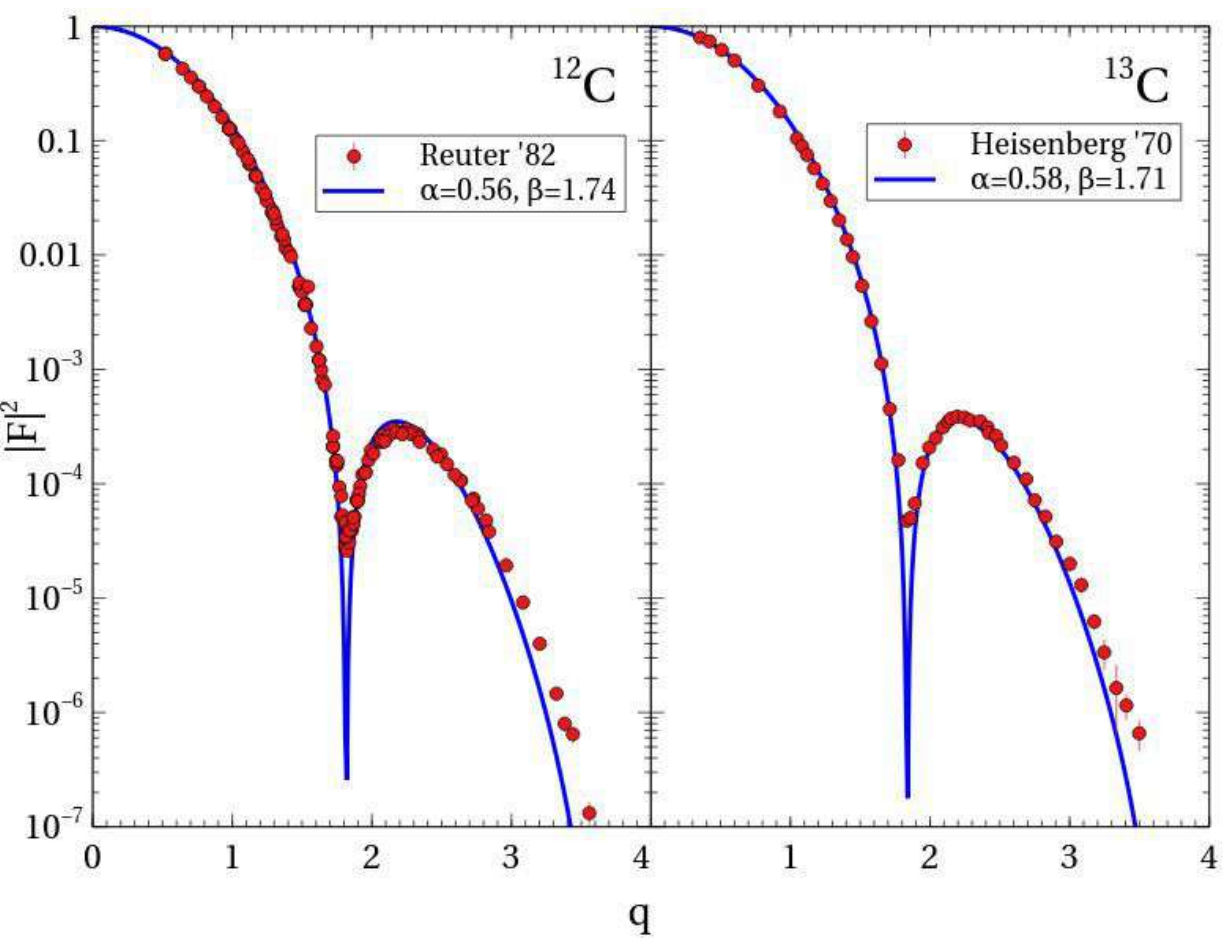}%
\caption{\label{FFElasComp}Comparison of elastic form factor between $^{12}$C (left) and $^{13}$C(right). Both graphs include available experimental data for both $^{12}$C\cite{reuter} and $^{13}$C\cite{elff}.}
\end{figure}

\begin{figure}[h]
\centering
\includegraphics[width=19pc,height=16pc]{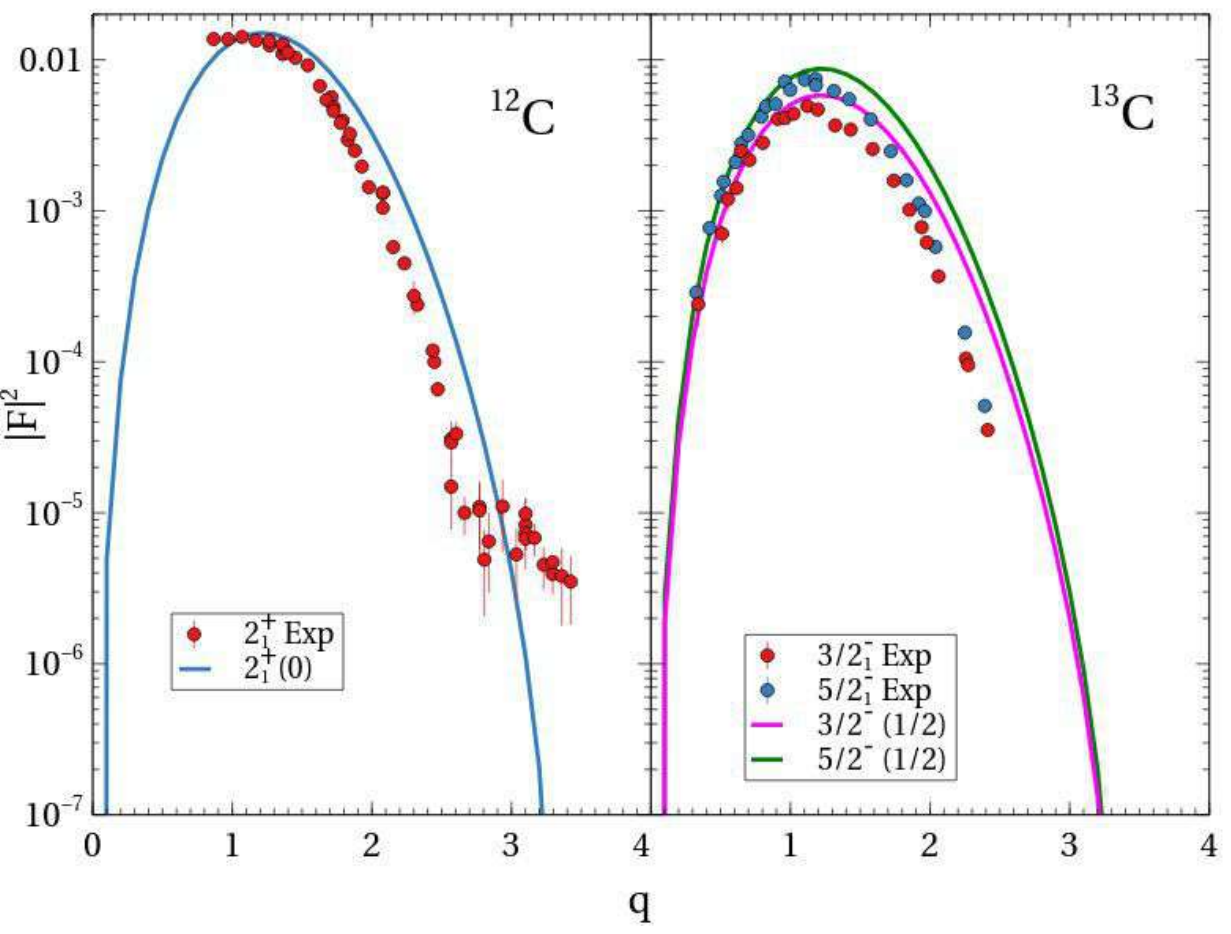}
\caption{\label{FFC2Comp}Comparison of C2 form factor between $^{12}$C (left) and $^{13}$C(right). Both graphs include available experimental data for both $^{12}$C\cite{crannell1,crannell2} and $^{13}$C\cite{inelff}.}
\end{figure}

A final remark arises when calculating form factors with diagonal intrinsic states. In that case, branching ratios are found between different form factors with the same multipole.
\begin{align}
\frac{\left|F_{C\lambda}(q;\Omega\gamma,I'K'\rightarrow\Omega\gamma,I_{1}K_{1})\right|^{2}}{\left|F_{C\lambda}(q;\Omega\gamma,I'K'\rightarrow\Omega\gamma,I_{2}K_{2})\right|^{2}}=\frac{\left<I',K',\lambda,K_{1}-K'|I_{1},K_{1}\right>^{2}}{\left<I',K',\lambda,K_{2}-K'|I_{2},K_{2}\right>^{2}}.\label{alagarule}
\end{align}
Finally one sees that the single particle element for Coulomb form factors doesn't give any contribution to the total value, thus we expect that for both $^{12}$C and $^{13}$C to have the same q dependence. The results are illustrated in \cref{FFElasComp,FFC2Comp}.

 \subsection{Transverse form factors}

 Just as previously done for the Coulomb form factors, the transverse form factors are obtained from a Fourier transform of the current $\vec{J}$ and magnetization operator $\vec{\mu}$, with a little more algebra involved to obtain the final expressions. The main assumption we do from the CSM is that the only contributor to both is the extra nucleon 
 \begin{align}
    \vec{J}&=0 \label{assumpTrans1}
    \\
    \vec{\mu}&=\delta\left(\vec{r_{i}}-\vec{x}\right)\mu_{i}\frac{\Vec{\sigma}\left(i\right)}{2M_{i}},\label{assumpTrans2}
 \end{align}
 where $\mu_{i}$ is the magnetic moment (in nuclear magnetons) of the nucleon involved, in this case for the free neutron is $\mu_{n}=-1.91$ and $M_{i}$ is the mass. 
 The multipoles for the transverse form factor are then given by
 \begin{align}
F_{E\lambda/M\lambda}(q;\Omega'\gamma',I'K'&=\left<I',K',\lambda,K-K'|I,K\right>\delta_{vv'}G_{\Omega\gamma,\Omega'\gamma'}^{{\rm sp}}(q) \nonumber
\\
&+P(-1)^{I+K}\left<I',K',\lambda,-K-K'|I,-K\right>\delta_{vv'}H_{\Omega\gamma,\Omega'\gamma'}^{{\rm sp}}(q),
\label{FFTlambda} 
 \end{align}
 where $G_{\Omega\gamma,\Omega'\gamma'}^{{\rm sp}}(q)$ and $H_{\Omega\gamma,\Omega'\gamma'}^{{\rm sp}}(q)$ will have the form
\begin{subequations}
\begin{align}
G^{\rm sp}_{\Omega\gamma,\Omega'\gamma'}(q) &=\sum_{nljm} \sum_{n'l'j'm'}  C^{\Omega\gamma}_{nljm} C^{\Omega'\gamma'}_{n'l'j'm'}\left< nljm \left| \hat{T}_{\lambda\mu}^{\zeta}\left(q,\vec{r}\,\right) \right| n'l'j'm' \right>\label{Gsp}
\\
H^{\rm sp}_{\Omega\gamma,\Omega'\gamma'}(q) &=\sum_{nljm} \sum_{n'l'j'm'}  C^{\Omega\gamma}_{nljm} C^{\Omega'\gamma'}_{n'l'j'm'} (-1)^{n+j+m}\left< nlj,-m \left| \hat{T}_{\lambda\mu}^{\zeta}\left(q,\vec{r}\,\right) \right| n'l'j'm' \right>\label{Hsp}
\end{align}
\label{ghsp}
\end{subequations}
and $\hat{T}_{\lambda\mu}^{\zeta}\left(q,\vec{r}\,\right)$ are the electric/magnectic transverse form factor operators
 \begin{align}
   \hat{T}_{\lambda\mu}^{E}\left(q,\vec{r}\,\right)&=q\frac{\mu_{i}}{2M_{N}}\vec{M}_{\lambda,\lambda}^{\mu}(\vec{r_{i}})\cdot\vec{\sigma}(i)\nonumber
   \\
   \hat{T}_{\lambda\mu}^{M}\left(q,\vec{r}\,\right)&=iq\frac{\mu_{i}}{2M_{N}}\left(-\sqrt{\frac{\lambda}{2\lambda+1}}\vec{M}_{\lambda,\lambda+1}^{\mu}(\vec{r_{i}})+\sqrt{\frac{\lambda+1}{2\lambda+1}}\vec{M}_{\lambda,\lambda-1}^{\mu}(\vec{r_{i}})\right)\cdot\vec{\sigma}(i) \nonumber
   \\
   \vec{M}_{\lambda_{1},\lambda_{2}}^{\mu}\left(\vec{r}\,\right)&=j_{\lambda_{2}}(qr)\vec{Y}_{\lambda_{1}\lambda_{2}1}^{\mu}(\theta,\phi)\label{OperadorMT}.
 \end{align}

\begin{figure}[h]
\centering
\includegraphics[width=16pc]{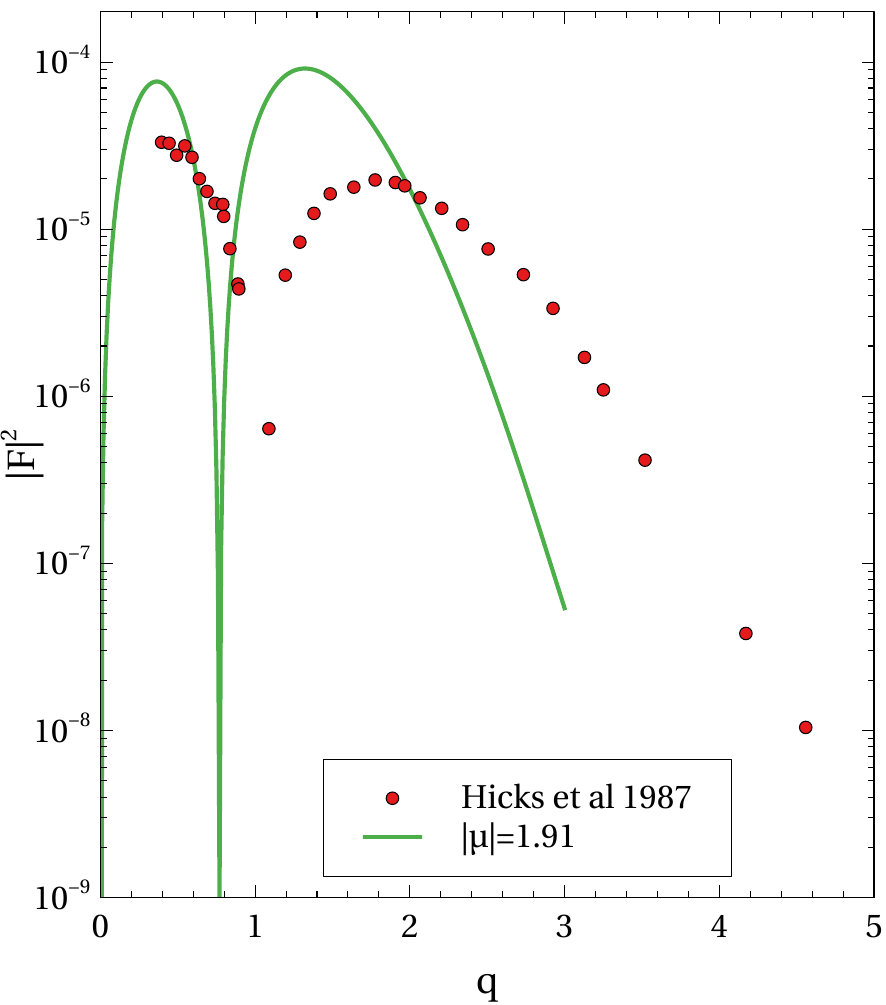}
\caption{\label{FFM1Elas}Comparison of M1 elastic form factor between experimental data \cite{inelff} and CSM calculation.}
\end{figure}

The results are showed in \cref{FFM1Elas}.

\section{Summary and Conclusions}
From calculations, it is seen that for longitudinal form factors in the base rotational band of $^{13}$C, the cluster core structure is dominant. In those cases, we expected that Coulomb form factors for both $^{12}$C and $^{13}$C have the same $q$ dependence. From \cref{FFElasComp,FFC2Comp} we see that the experimental data seems to support that idea, differing only slightly in some aspects. Also, it shows for $^{13}$C that longitudinal form factors follow a branching ratio rule like the one in \cref{alagarule}. This is of great importance since it indicates that some properties of the cluster structure in $^{12}$C are still present in $^{13}$C. 

Unfortunately, such a positive outcome was not the case for the transverse form factors. From \cref{FFM1Elas} we are not able to reproduce available experimental data. Only global functional properties are reproduced of \cite{inelff}. It is most likely that our assumptions in \cref{assumpTrans1,assumpTrans2} are incorrect and must then include a contribution of the cluster core. A possible solution may be found in \cite{DelormeFigurau}, but it would have to be adapted for the CSM.

To reiterate, the CSM as is now is a good approximation to calculate longitudinal form factors for $^{13}$C, reproducing accurately the available experimental data. Meanwhile, in the case of transverse form factors, the model is still lacking, due most likely to not considering the contribution of the cluster structure to the transverse form factors. 

\ack 
This work was supported in part by research grants IN101320, IG101423 from PAPIIT-DGAPA and 784896 from CONACYT.

\section*{References}

\bibliography{CocoyocAdrian}

\end{document}